\documentclass[aps,prl,reprint,superscriptaddress]{revtex4-2}
\usepackage[dvipsnames]{xcolor}
\usepackage{physics}
\usepackage{braket}
\usepackage{scalerel}
\usepackage{blindtext} 
\usepackage{epsfig, cancel}

\usepackage{latexsym}
\usepackage{natbib, comment}
\usepackage{mathrsfs,amsmath,amssymb,amsthm,amsfonts,tikz,graphicx,accents,hyperref,color}
\usepackage{url}
\usepackage{dcolumn}
\usepackage{multirow}
\usepackage{color}
\usepackage{cancel}
\usepackage{soul}
\usepackage[normalem]{ulem}
\usepackage{txfonts}
\usepackage{epsfig}
\usepackage{psfrag}
\usepackage{subfigure}
\hypersetup{colorlinks=true}
\usepackage{mathtools}
\usepackage{enumitem}
\usepackage{float}
\usepackage{caption,ragged2e}
\usetikzlibrary{decorations.markings,fpu}

\usetikzlibrary{backgrounds}

\usetikzlibrary{calc}

\def\H0{{\text{H}\hspace*{-2.05mm}\text{H} 0\hspace*{-1.35mm}0\ }}

\usepackage[all]{xy}

\usepackage{slashed}
\usepackage{slashed,ccaption}
\usepackage{multirow}
\usetikzlibrary{calc} 

\usepackage{caption}

\usepackage{array}
\hypersetup{ linktoc=all,
    colorlinks, linkcolor={palatinateblue},
    citecolor={red}, urlcolor={amaranth} 
}

\graphicspath{{Images/}}

\DeclareSymbolFont{extraup}{U}{zavm}{m}{n}
\DeclareMathSymbol{\varheart}{\mathalpha}{extraup}{86}
\DeclareMathSymbol{\vardiamond}{\mathalpha}{extraup}{87}
\makeatletter
\renewcommand*{\@fnsymbol}[1]{\ensuremath{\ifcase#1\or \clubsuit \or \vardiamond \or \varheart\or
    \spadesuit\or \mathparagraph\or \|\or **\or \dagger\dagger
    \or \ddagger\ddagger \else\@ctrerr\fi}}
\makeatother

\definecolor{rosy}{RGB}{230,235,252}
\definecolor{myframetitle}{RGB}{90,89,170}
\definecolor{myblocktitle}{RGB}{140,185,249}
\definecolor{mytitle}{RGB}{10,80,26}

\definecolor{darkgreen}{RGB}{27,130,45}
\definecolor{darkblue}{rgb}{0,0,0.3}
\definecolor{darkred}{rgb}{0.7,0,0}

\definecolor{light gray}{RGB}{220,220,220}
\definecolor{dark purple}{RGB}{108,0,217}
\definecolor{pink}{RGB}{190,20,100}
\definecolor{orang}{RGB}{193,63,0}
\definecolor{green}{RGB}{11,98,17}
\definecolor{darkpink}{RGB}{153,0,76}
\definecolor{bluegreen}{RGB}{0,102,102}
\definecolor{greenlagan}{RGB}{0,102,0}
\definecolor{redgreen}{RGB}{102,102,0}
\definecolor{Redgreen}{RGB}{153,76,0}
\definecolor{vividviolet}{rgb}{0.62, 0.0, 1.0}
\definecolor{amaranth}{rgb}{0.9, 0.17, 0.31}
\definecolor{palatinateblue}{rgb}{0.15, 0.23, 0.89}
\definecolor{brightpink}{rgb}{1.0, 0.0, 0.5}
\definecolor{cornflowerblue}{rgb}{0.39, 0.58, 0.93}
\definecolor{deepcarminepink}{rgb}{0.94, 0.19, 0.22}
\definecolor{radicalred}{rgb}{1.0, 0.21, 0.37}

\newcommand\ignore[1]{}
\usepackage[most]{tcolorbox}

\tcbset{highlight math style={left=02mm,right=02mm,top=02mm,bottom=02mm}} 
\usepackage{empheq}

\newcommand\inbox[1]{\tcbset{fonttitle=\scriptsize} \tcboxmath[colback=white,colframe=black!70]{#1}}

\begin{document}

\newcommand{\mytitle}{{\large{\textbf{Gravity/Thermodynamics Correspondence for Generic Dynamical Spacetimes}}}}
\title{{\mytitle}}

\author{V.~R.~Shajiee}\email{v.shajiee@ipm.ir}
\affiliation{School of Physics, Institute for Research in Fundamental Sciences (IPM), P.O.Box 19395-5531, Tehran, Iran}

\author{M.~M.~Sheikh-Jabbari}\email{jabbari@theory.ipm.ac.ir}
\affiliation{School of Physics, Institute for Research in Fundamental Sciences (IPM), P.O.Box 19395-5531, Tehran, Iran}

\author{V.~Taghiloo}\email{v.taghiloo@iasbs.ac.ir}
\affiliation{Department of Physics, Institute for Advanced Studies in Basic Sciences (IASBS), P.O.Box 45137-66731, Zanjan, Iran}
\affiliation{School of Physics, Institute for Research in Fundamental Sciences (IPM), P.O.Box 19395-5531, Tehran, Iran}

\begin{abstract}
Working within covariant phase space formalism, we establish \textit{gravity-thermodynamics correspondence} for generic dynamical spacetimes. We derive local first and second laws of thermodynamics on general null boundaries, for generic non-stationary solutions to Einstein gravity with minimally coupled matter for physical observers that we carefully specify. Entropy variation, which receives contribution only from the gravitational sector, is the surface charge variation associated with a vector field generating the null boundary, and temperature is its non-affinity parameter (divided by $2\pi$). The statement of the second law in our setting is that the entropy variation evaluated along the null vector field is strictly non-negative for the admissible positive-non-affinity clock and matter satisfying the null energy condition. Our dynamical entropy variation, which is non-integrable for generic non-stationary solutions, becomes integrable in near-stationary configurations, reproducing the Hollands--Wald--Zhang dynamical entropy formula. Strikingly, this local second law holds even when the boundary area is decreasing,  disentangling thermodynamic irreversibility and the second law from standard area theorems. Our first-principles construction demonstrates that gravitational thermodynamics requires neither a stationary background, teleological event horizons, nor an exact entropy state function on individual codimension-1 spacelike slices of the null boundary (codimension two in spacetime).
\end{abstract}
\maketitle

\section{Introduction}
\label{sec:introduction}

The relation between gravity, as described by General Relativity (GR), and thermodynamics has been studied for over half a century, initially started by exploring black holes \cite{Bekenstein:1972tm, Bardeen:1973gs, Hawking:1974sw}. However, this relation was promoted to something much deeper and more profound, to a correspondence/equivalence, by Jacobson's observation that the Einstein field equations function from the Clausius relation \cite{Jacobson:1995ab}; see also \cite{Padmanabhan:2003gd, Kolekar:2011gw}. This gravity/thermodynamic correspondence posits that thermal properties are not only a property of event horizons or stationary black holes, but are an inherent consequence of diffeomorphism invariance, a manifestation of (Einstein's) equivalence principle, whenever the observer finds a null boundary, in accord with the observer-dependent Unruh effect \cite{Unruh:1976db}. 

The observers confined to one side of the null boundaries do not have causal access to the information on the other side. The picture that has emerged is that local thermodynamic relations at the null boundary substitute for information about the gravitational dynamics not accessible to the observer, as necessitated by the equivalence principle. Our analyses and discussions in this paper is to clarify and provide further support for this picture, extending and generalizing analyses in \cite{Jacobson:1995ab}.

In most of the literature with regard to thermodynamics in gravitating systems, the main focus has been on the spacetime and its properties and features, typically black holes and their horizons. In this work, following the viewpoints discussed in \cite{Golshani:2024fry} and further developed in \cite{Golshani:2026lvc}, we note that besides the spacetime, one may focus on the solution space, the space of all physically distinct solutions in GR with prescribed {fall-off} conditions; each point in the solution space represents an exact, fully non-linear solution to the field equations. As highlighted in \cite{Golshani:2024fry}, this solution space framework is connected with the background independence in GR descriptions, while the spacetime picture is a manifestation of the general covariance.

The appropriate framework to study the solution space is the covariant phase space formalism (CPSF) \cite{Lee:1990nz, Iyer:1994ys}. In this framework, a generic solution corresponds to a generic configuration of the associated thermodynamic system which can be far from equilibrium. The solution phase-space variation $\delta$ is the difference of two nearby points in the solution space and hence satisfies the coupled linearized field equations. While most of the literature concerns thermodynamical description of (near) stationary gravitational solutions, extending the description to generic non-stationary situations has also been discussed in e.g.~\cite{Ashtekar:2002ag, Ashtekar:2003hk, Hayward:2004fz, Eling:2006aw, Chirco:2009dc, Freidel:2013jfa, Freidel:2014qya, Adami:2021kvx, Ashtekar:2025qqa}.

The solution space viewpoint is more naturally connected to the thermodynamic description, where one deals with a thermodynamic configuration space. Formulating a rigorous thermodynamic description in a generic non-stationary regime requires three essential ingredients: a generic solution of the underlying theory, an arbitrary null boundary, and a specified physical observer (or clock) to measure local thermodynamic quantities. Previous studies have constructed the phase space and surface charges of generic null boundaries in Einstein gravity~\cite{Adami:2021nnf, Hopfmuller:2016scf, Chandrasekaran:2018aop, Chandrasekaran:2021hxc, Odak:2023pga, Donnay:2015abr} and formulated a local first law for the associated open thermodynamic systems~\cite{Adami:2021kvx, Freidel:2013jfa}. To perform the analyses, we employ CPSF with an emphasis on the geometric aspects of the solution phase space, including surface charge (non)integrability, change of slicing, and genuine geometric and physical integrability obstructions associated with gravitational news~\cite{Golshani:2026lvc}. 

In this paper, we complete the gravity/thermodynamics correspondence for generic dynamical spacetimes. Working in Einstein gravity on a spacetime with an arbitrary null boundary, we introduce an admissible physical clock to evaluate the covariant surface charge variations. This establishes a well-defined notion of the temperature as well as an exact local first law, allowing us to isolate the heat sector and extract a dynamical entropy one-form on the solution space.

A central consequence of our exact formulation is that this entropy variation is generically non-integrable, rendering the accumulated entropy fundamentally history-dependent. However, near a stationary solution, it becomes integrable, matching the ``dynamical entropy'' recently derived by Hollands, Wald, and Zhang (HWZ) \cite{Hollands:2024vbe}. (See also the recent developments on this subject \cite{Visser:2024pwz, Rignon-Bret:2023fjq, Bhattacharyya:2026aiu, Imseis:2026zgf, Chakraborty:2026emn, Jia:2025tgf}.) The physical identification of this entropy variation is ultimately confirmed by its adherence to a local second law with a matter content that satisfies the null energy condition (NEC). 

Without restricting to stationary backgrounds, invoking teleological future endpoints, or relying on perturbative expansions, we prove that, under the NEC for the matter content {and positive clock non-affinity}, this dynamical entropy satisfies an exact, non-perturbative local second law, where local entropy production is strictly non-negative and driven entirely by the gravitational shear squared plus the Ricci tensor contracted with the null vector field generating the null boundary.

\vspace*{-2mm}
\section{Geometric setup and admissible variations}
\label{subsec:geometry-variations}
\vspace*{-2mm}

Consider Einstein gravity in a $d$-dimensional spacetime $\mathcal{M}$ admitting a smooth null boundary $\mathcal N$ foliated by spacelike codimension-2 slices $\Sigma$. Let $l^\mu$ denote a future-directed null generator of $\mathcal N$ and $n^\mu$ the auxiliary null normal to $\Sigma$, normalized such that $l\cdot n=-1$. Extending this frame smoothly off the boundary, the metric decomposes as
\begin{equation}
 g_{\mu\nu}=-2l_{(\mu} n_{\nu)}+q_{\mu\nu},
 \qquad
 q_{\mu\nu}l^\nu=q_{\mu\nu}n^\nu=0,
 \label{eq:ngc-metric}
\end{equation}
where $q_{\mu\nu}$ induces a positive-definite metric on each cut $\Sigma$. We define the binormal $\epsilon_{\mu\nu}=2l_{[\mu}n_{\nu]}$.

The kinematics of the null generators are captured by the deformation tensor $\Theta^l_{\mu\nu}$, which further splits into the expansion $\theta_l$ and shear $N^l_{\mu\nu}$:
\begin{equation}\begin{split}
 \Theta^l_{\mu\nu}
 :=q_\mu{}^\rho q_\nu{}^\sigma\nabla_\rho l_\sigma,
 &\qquad \theta_l:=q^{\mu\nu}\Theta^l_{\mu\nu},
\\
 N^l_{\mu\nu}
 :=\Theta^l_{\mu\nu}-\frac{\theta_l}{d-2}q_{\mu\nu},&\qquad l^\nu\nabla_\nu l^\mu :\mathrel{\hat{=}}\kappa_l l^\mu,
 \label{eq:ngc-deformation}
\end{split}\end{equation}
where $\mathrel{\hat{=}}$ denotes equality restricted to $\mathcal N$ and 
the non-affinity $\kappa_l$ depends on the chosen normalization of $l$. Hypersurface orthogonality guarantees that $\Theta^l_{\mu\nu}$ is symmetric. The $d-2$ area-form on $\Sigma$,  $\boldsymbol\epsilon_\Sigma$, evolves along the generators as
\begin{equation}
 \mathcal L_l\boldsymbol\epsilon_\Sigma
   =\theta_l\boldsymbol\epsilon_\Sigma,
 \qquad
 \delta\boldsymbol\epsilon_\Sigma
   =\frac12\boldsymbol\epsilon_\Sigma q^{\mu\nu}\delta q_{\mu\nu},
 \label{eq:ngc-area}
\end{equation}
where $\mathcal L_l$ denotes Lie derivative along $l^\mu$.

To compare solutions while holding the boundary ${\cal N}$ and the slice $\Sigma$ on it fixed, we introduce a boundary-adapted phase space with the co-normal variations satisfying \cite{Parattu:2015gga},
\begin{equation}
 \delta l_\mu\mathrel{\hat{=}} \delta \phi\, l_\mu,
 \qquad
 \delta n_\mu\mathrel{\hat{=}}\delta \psi\, n_\mu.
 \label{eq:ngc-admissible-variations}
\end{equation}
While these conditions fix the geometric directions of the null covector variations over the solution space, they permit the intrinsic geometry of the cut $\Sigma$ and the normalizations of the null vectors to remain fully dynamical. This relative shift in normalizations can be tracked by introducing the boost-variation parameter $b[\delta]$, which is a one-form over the solution space:
\begin{equation}
\begin{split}
 b[\delta]
 &:=-n^\mu\delta l_\mu+l^\mu\delta n_\mu
   =\delta(\phi-\psi),
\\
 \mathfrak D_\delta\theta_l
 &:=\delta\theta_l-\frac12\theta_l b[\delta].
 \label{eq:ngc-normal-variation}
\end{split}
\end{equation}
$\mathfrak D_\delta\theta_l$ may be viewed as covariant derivative over the solution space of the theory.
Under a local boost,
\begin{equation}\label{boost-l,n}
    l\mapsto e^\sigma l, \qquad n\mapsto e^{-\sigma}n,
\end{equation}
hence $g_{\mu\nu}, q_{\mu\nu}$ are invariant under local boost and,
\begin{equation}\label{boost-b,theta}
\kappa_l\mapsto e^\sigma(\kappa_l+{\cal L}_l\sigma),\quad    b\mapsto b+2\delta\sigma,\quad \mathfrak D_\delta\theta_l\mapsto e^\sigma\mathfrak D_\delta\theta_l. 
\end{equation}
Finally, rotational and tangential features are sourced by the tangential projection of the Lie bracket of $n^\mu$ and $l^\mu$, 
\begin{equation}
 J_\mu=
    q_\mu{}^\nu (l^\rho\nabla_\rho n_\nu- n^\rho\nabla_\rho l_\nu)\,.
 \label{eq:ngc-tangential-momentum}
\end{equation}
$J_\mu$ is invariant under boosts. All quantities are defined geometrically, avoiding reliance on any specific coordinate system. We stress that the choices made in this part impose no restriction on the spacetime geometry.

\section{Surface-charge variations}
\label{subsec:symplectic-charges}

We develop our formulation within the CPSF worked out for the Einstein--Hilbert theory  with minimally coupled matter, described by the $d$-form Lagrangian \cite{Iyer:1994ys},
\begin{equation}
  \boldsymbol L =\frac{R-2\Lambda}{16\pi G}\boldsymbol\epsilon {+ \boldsymbol L_{\mathrm m}(g,\Psi,\nabla\Psi)}\, ,
\end{equation}
where $\boldsymbol\epsilon$ is the volume form of $\mathcal{M}$, $\Psi$ denotes the matter fields, and $\Phi=(g,\Psi)$. We assume a diffeomorphism-covariant matter Lagrangian $\boldsymbol L_{\mathrm m}$ with no explicit curvature coupling. Its variation with $\delta g_{\mu\nu}:=h_{\mu\nu}$ defines the stress tensor and matter symplectic potential by
\begin{align}
 \delta\boldsymbol L_{\mathrm m}
 =&\frac12 T^{\mu\nu}h_{\mu\nu}\boldsymbol\epsilon
   +\boldsymbol E_\Psi\cdot\delta\Psi + \mathrm d\boldsymbol\Theta_{\mathrm m}(\Phi;\delta\Phi),
 \label{eq:matter-variation}\\
 G_{\mu\nu}&+\Lambda g_{\mu\nu}=8\pi G T_{\mu\nu},
 \qquad \boldsymbol E_\Psi=0.
 \label{eq:coupled-field-equations}
\end{align}
$\boldsymbol L,\boldsymbol\Theta,\boldsymbol Q_\xi,\boldsymbol k_\xi$, and $H_\xi$ below refer to the full (gravity + matter) theory, while the same quantities with subscript ${\mathrm m}$ denotes the matter contribution. We allow unrestricted smooth matter data on the null boundary and variations satisfying the coupled linearized equations. If further matter boundary conditions are imposed, both the variations and physical evolution must preserve them. No vanishing matter flux or matter surface-charge variation is imposed. We use the diffeomorphism action $\delta_\xi\Psi=\mathcal L_\xi\Psi$.
 The standard Lee--Wald symplectic potential~\cite{Lee:1990nz, Iyer:1994ys} for a generic {field variation $\delta\Phi=(h,\delta\Psi)$} is
\begin{equation}
 \begin{aligned}
 \boldsymbol\Theta({\Phi;\delta\Phi})
 &=\frac{\boldsymbol\epsilon_\mu}{16\pi G}
       \left(\nabla_\nu h^{\mu\nu}-\nabla^\mu h\right) {+ \boldsymbol\Theta_{\mathrm m}(\Phi;\delta\Phi)},
 \end{aligned}
 \label{eq:ngc-lagrangian-potential}    
\end{equation}
where $\boldsymbol\epsilon_\mu$ denotes {the contracted spacetime volume form, $(\boldsymbol\epsilon_\mu)_{\mu_2\cdots\mu_d}=\epsilon_{\mu\mu_2\cdots\mu_d}$}. We use $h^{\mu\nu}=g^{\mu\rho}g^{\nu\sigma}h_{\rho\sigma}$ and $h=g^{\mu\nu}h_{\mu\nu}$.

For a vector field $\xi^\mu$ which may  depend on solution space fields ($\delta \xi\neq 0$), the associated  Noether $(d-2)$-form charge density $\boldsymbol Q_\xi$ and the covariant phase space surface charge density $\boldsymbol k_\xi$ and corresponding charge variation $\slashed{\delta} H_\xi$ are,
\begin{align}
(\boldsymbol Q_\xi)_{\mu_1\cdots\mu_{d-2}}
 &:=-\frac{1}{16\pi G}
   \epsilon_{\mu_1\cdots\mu_{d-2}\rho\sigma}
   \nabla^\rho\xi^\sigma {+ (\boldsymbol Q_\xi^{\mathrm m})_{\mu_1\cdots\mu_{d-2}}}\,,
 \label{eq:ngc-noether-definition} \\
 \boldsymbol k_\xi[ {\Phi;\delta\Phi}]
 &:=\delta\boldsymbol Q_\xi-\boldsymbol Q_{\delta\xi}
                     -\iota_\xi\boldsymbol\Theta({\Phi;\delta\Phi}),
\\
 \slashed{\delta} H_\xi
 &:=\int_\Sigma\boldsymbol k_\xi[{\Phi;\delta\Phi}].
 \label{eq:ngc-charge-prescription}
\end{align}
{The matter contribution $\boldsymbol Q_\xi^{\mathrm m}$ is obtained from the diffeomorphism Noether current of $\boldsymbol L_{\mathrm m}$, with similar freedoms/ambiguity prescription as in the gravitational sector. In particular,}
{
\begin{align}
 \boldsymbol k_\xi^{\mathrm m}
 &:=\delta\boldsymbol Q_\xi^{\mathrm m}
    -\boldsymbol Q_{\delta\xi}^{\mathrm m}
    -\iota_\xi\boldsymbol\Theta_{\mathrm m},
 \label{eq:matter-surface-charge}\\
 \slashed{\delta}H_\xi^{\mathrm m}
 &:=\int_\Sigma\boldsymbol k_\xi^{\mathrm m},
 \qquad
 \boldsymbol k_\xi=\boldsymbol k_\xi^{\mathrm g}+\boldsymbol k_\xi^{\mathrm m}.
 \label{eq:matter-charge-variation}
\end{align}}
{The matter term is retained explicitly because its reduction to particular work or chemical-potential terms depends on the matter Lagrangian and boundary conditions.}

Note that the Noether charge density $\boldsymbol Q_\xi$ is a scalar {on the solution space}, whereas $\boldsymbol k_\xi[{\Phi;\delta\Phi}]$ and $\slashed{\delta} H_\xi$ are one-forms on the solution space. These charges are defined up to well-known and understood freedoms (ambiguities) \cite{Iyer:1994ys, Golshani:2026lvc}. One of the freedoms, the $W$-freedom, which is associated with a shift of the Lagrangian by a total derivative (boundary) term, only affects the Noether charge, and the other one, the $Y$-freedom, which affects both $\boldsymbol Q_\xi, \slashed{\delta} H_\xi$, is fixed by setting the corner terms to zero; see \cite{Golshani:2026lvc} for more detailed discussions. The presymplectic current $\boldsymbol\omega:=\delta\boldsymbol\Theta$ {of the full theory, evaluated on coupled solutions and variations satisfying the coupled linearized equations,}  satisfies the fundamental identity \cite{Lee:1990nz, Iyer:1994ys, Wald:1999wa}
\begin{equation}
 \mathrm d\boldsymbol k_\xi[{\Phi;\delta\Phi}]
   =\boldsymbol\omega({\Phi;\delta\Phi,\mathcal L_\xi\Phi}),
 \label{eq:ngc-charge-identity}
\end{equation}
naturally accommodating symplectic fluxes across $\mathcal N$ due to the dynamical metric over $\Sigma$, $q_{\mu\nu}$ and the matter fields.

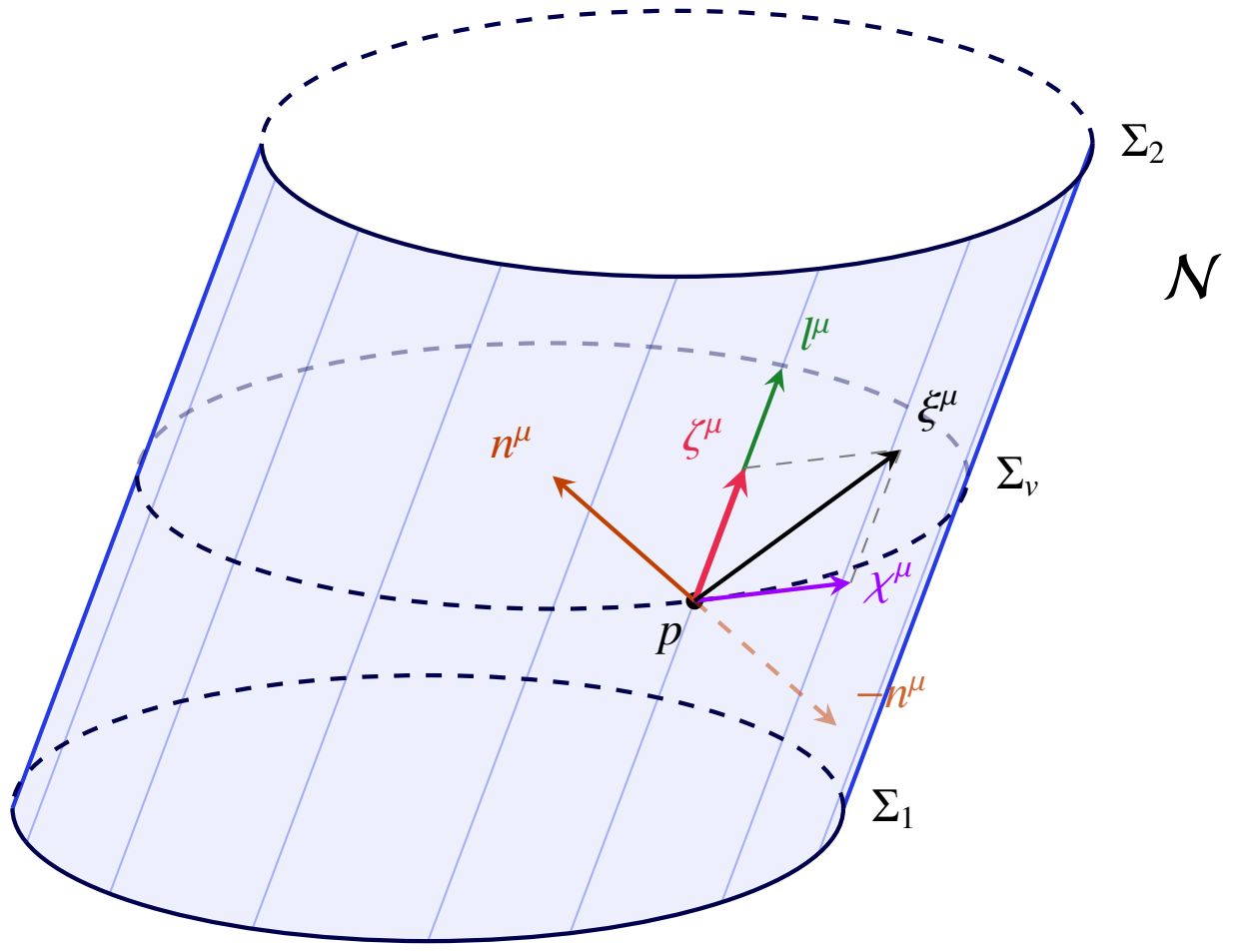
\begin{figure}[t]
\centering
\begin{tikzpicture}[scale=1.1, >=stealth]
    
    \colorlet{Nfill}{palatinateblue!8}
    \colorlet{Nedge}{palatinateblue}
    \colorlet{CutColor}{darkblue}

    \def\dx{1.5}
    \def\dy{4.0}

    \def\rx{2.5}
    \def\ry{0.8}

    \fill[Nfill] (-\rx, 0) arc (180:360:\rx\space and \ry) -- 
                 (\dx+\rx, \dy) arc (360:180:\rx\space and \ry) -- cycle;

    \draw[thick, Nedge] (-\rx, 0) -- (\dx-\rx, \dy);
    \draw[thick, Nedge] (\rx, 0) -- (\dx+\rx, \dy);

    \draw[palatinateblue!40, thin] (-2.415, -0.207) -- ++(\dx, \dy); 
    \draw[palatinateblue!40, thin] (-1.915, -0.514) -- ++(\dx, \dy); 
    \draw[palatinateblue!40, thin] (-1.056, -0.725) -- ++(\dx, \dy); 
    \draw[palatinateblue!40, thin] (0.000, -0.800) -- ++(\dx, \dy);  
    \draw[palatinateblue!40, thin] (0.855, -0.752) -- ++(\dx, \dy);  
    \draw[palatinateblue!40, thin] (1.768, -0.566) -- ++(\dx, \dy);  
    \draw[palatinateblue!40, thin] (2.349, -0.274) -- ++(\dx, \dy);  

    \node[font=\large] at (\dx+\rx+0.6, \dy-0.8) {$\mathcal{N}$};

    \draw[thick, CutColor, dashed] (\rx, 0) arc (0:180:\rx\space and \ry);
    \draw[thick, CutColor] (-\rx, 0) arc (180:360:\rx\space and \ry) node[right, black, xshift=2pt] {$\Sigma_1$};

    \draw[thick, CutColor, dashed] (\dx+\rx, \dy) arc (0:180:\rx\space and \ry);
    \draw[thick, CutColor] (\dx-\rx, \dy) arc (180:360:\rx\space and \ry) node[right, black, xshift=2pt] {$\Sigma_2$};

    \draw[thick, CutColor, dashed, opacity=0.4] (0.5*\dx+\rx, 0.5*\dy) arc (0:180:\rx\space and \ry);
    \draw[thick, CutColor, dashed] (0.5*\dx-\rx, 0.5*\dy) arc (180:360:\rx\space and \ry) node[right, black, xshift=2pt] {$\Sigma_v$};

    \coordinate (P) at (1.605, 1.248);
    \fill (P) circle (1.5pt) node[below left, yshift=-1pt, xshift=1pt] {$p$};

    \path (P) ++(0.052, 0.140) coordinate (C1);
    \path (P) ++(-0.033, 0.215) coordinate (C2);
    \path (P) ++(-0.085, 0.075) coordinate (C3);

    \draw[->, thick, darkgreen] (P) -- ++(0.525, 1.400) node[above right, xshift=0pt] {$l^\mu$};

    \draw[->, thick, orang] (P) -- ++(-0.855, 0.752) node[above left] {$n^\mu$};

    \draw[->, thick, dashed, orang, opacity=0.5] (P) -- ++(0.855, -0.752) node[above right, opacity=0.8] {$-n^\mu$};

    \draw[->, very thick, amaranth] (P) -- ++(0.300, 0.800) node[above left, yshift=-2pt] {$\zeta^\mu$};

    \draw[->, thick, vividviolet] (P) -- ++(0.940, 0.110) node[right] {$\chi^\mu$};

    \draw[->, thick, black] (P) -- ++(1.240, 0.910) node[above right] {$\xi^\mu$};
    
    \draw[dashed, thin, gray] (P) ++(0.300, 0.800) -- ++(0.940, 0.110);
    \draw[dashed, thin, gray] (P) ++(0.940, 0.110) -- ++(0.300, 0.800);

\end{tikzpicture}
\caption{\justifying
Geometric decomposition on a dynamical null boundary $\mathcal{N}$ foliated by spacelike sections $\Sigma_v$. The null generator $l^\mu$ and the transverse auxiliary normal $n^\mu$ are normalized such that $l \cdot n = -1$. The prescribed physical clock $\zeta^\mu$ lies strictly along the boundary generator ($\zeta^\mu \mathrel{\hat{=}} f l^\mu$). A generic canonical evolution vector $\xi^\mu = \zeta^\mu + \chi^\mu$ incorporates a spatial shift $\chi^\mu$ tangent to $\Sigma_v$, identifying the canonical source of tangential {angular-}momentum work in the exact local first law. To visually distinguish the normal plane, $n^\mu$ is shown pointing radially inward, while $-n^\mu$ extends symmetrically outward. {The directions are schematic; $l\cdot n=-1$ does not denote orthogonality.}}
\label{fig:geometric_setup}
\end{figure}

\vskip 2mm

\textbf{Symmetry generators.} Near the boundary ${\cal N}$, we decompose the symmetry generator $\xi$ as,
\begin{equation}
 \xi^\mu=f l^\mu+Z n^\mu+\chi^\mu,
 \qquad q^\mu{}_{\nu}\chi^\nu=\chi^\mu.
 \label{eq:ngc-generator}
\end{equation}
Preserving the fixed null boundary requires $Z\mathrel{\hat{=}}0$. We restrict our attention to generators whose induced variations respect the conditions \eqref{eq:ngc-admissible-variations}. $\chi^\mu$ is invariant under boosts \eqref{boost-l,n} and we choose
\begin{equation}\label{boost-f,Z}
    f\mapsto e^{-\sigma} f,\qquad Z\mapsto e^\sigma Z,
\end{equation}
such that $\xi$ is boost invariant. Although $Z$ vanishes on ${\cal N}$, its transverse derivative does not.  Defining 
\begin{equation}\label{W-K-def}
   {\cal W}:=-n^\mu\nabla_\mu Z\big|_{\mathcal N},\qquad {\cal K}:=f\kappa_l+\mathcal L_l f+ {\cal W},
\end{equation} 
the pullback of the Noether charge on $\Sigma$ evaluates to
\begin{equation}
 \begin{aligned}
 \boldsymbol Q_\xi\big|_{_\Sigma}
 &=-\frac{\boldsymbol\epsilon_\Sigma}{16\pi G}
       \epsilon_{\mu\nu}\nabla^\mu\xi^\nu
       {+ \boldsymbol Q_\xi^{\mathrm m}\big|_\Sigma}\\
 &=\frac{\boldsymbol\epsilon_\Sigma}{16\pi G}
       \left({\cal K}+\chi^\mu J_\mu\right)
       + \boldsymbol Q_\xi^{\mathrm m}\big|_\Sigma.
 \end{aligned}
 \label{eq:ngc-noether-charge}
\end{equation}
{Consequently, the Noether charge is determined not only by the boundary value of $\xi$, but also by its transverse derivatives. Furthermore, applying the transformations \eqref{boost-l,n} and \eqref{boost-f,Z} reveals that $\mathcal{W}$, $\mathcal{K}$, and thus the entire Noether charge \eqref{eq:ngc-noether-charge}, are manifestly invariant under local boosts.}

Substituting the admissible variations~\eqref{eq:ngc-admissible-variations} and the generator decomposition~\eqref{eq:ngc-generator} into the charge prescription~\eqref{eq:ngc-charge-prescription}, we obtain the surface-charge variation over $\Sigma$:
\begin{align}
 \slashed{\delta} H_\xi
 = \slashed{\delta}H_\xi^{\mathrm m}+ \frac{1}{16\pi G}&\int_\Sigma
 \Big[{\cal K}\,\delta\boldsymbol\epsilon_\Sigma+\chi^\mu\delta\!\left(
                     \boldsymbol\epsilon_\Sigma J_\mu\right)
 \nonumber\\  &\quad
     -f\boldsymbol\epsilon_\Sigma\,\left(2
                     \mathfrak D_\delta\theta_l+
                     \Theta_l^{\mu\nu}\delta q_{\mu\nu} \right)
 \Big]
 \label{eq:ngc-full-charge}
\end{align}
Remarkably, the expansion and normal-normalization variations persist even though $Z$ vanishes identically on the boundary. $\slashed{\delta} H_\xi$ in \eqref{eq:ngc-full-charge} forms the basis for the thermodynamic interpretation developed below.

\section{Physical observers and the local temperature}
\label{subsec:evolution-generator}

The notion of energy is conjugate to a chosen notion of time evolution and is hence observer-dependent; so are {the temperature conjugate to the entropy variation, and the entropy and its variation}. Therefore, to discuss thermodynamic quantities (charges and associated chemical potentials) one needs to specify how a physical observer reports them.

To specify the thermodynamic symmetry generators and associated chemical potentials, consider a future-directed null direction on $\mathcal N$ and extend it smoothly to a local neighborhood:
\begin{equation}
 \zeta^\mu=f l^\mu+Z n^\mu,
 \qquad f>0,
 \qquad Z\mathrel{\hat{=}}0.
 \label{eq:elf-thermal-evolution}
\end{equation}
On the boundary ${\cal N}$, the non-affinity of this generator is
\begin{equation}
 \zeta^\nu\nabla_\nu\zeta^\mu
   \mathrel{\hat{=}}\kappa_\zeta\zeta^\mu,
 \qquad
 \kappa_\zeta=f\kappa_l+\mathcal L_l f.
 \label{eq:elf-nonaffinity}
\end{equation}
To ensure $\zeta$ represents an admissible clock, its extension must preserve the normal-covector conditions~\eqref{eq:ngc-admissible-variations}. Specifically, this requires
\begin{equation}
 \mathcal L_\zeta n_\mu
 \mathrel{\hat{=}}f\mathcal L_l n_\mu-\nabla_\mu f
 \mathrel{\hat{=}} \delta_\zeta \psi\, n_\mu.
 \label{eq:elf-admissible-clock}
\end{equation}
One must ensure that the above equation is integrable, i.e., that there exist choices of $f$ that make the left-hand side a vector proportional to $n_\mu$. 

The detailed analysis may be found in Appendix~\ref{app:admissible-clock}; here we present a sketch of the argument. First, one may observe that projecting both sides of \eqref{eq:elf-admissible-clock} along $l^\mu$,  fixes $\delta_\zeta\psi \mathrel{\hat{=}} \mathcal L_l f$, while projection along $n^\mu$ specifies  $n^\mu\partial_\mu f$ just off the boundary. Next, let $v$ label the spatial cuts on ${\cal N}$, $\Sigma_v$ (cf. Fig.\ref{fig:geometric_setup}), and define the geometric lapse $\tau := 1/(l^\mu \nabla_\mu v)$. The tangential equation integrates exactly to yield $f \mathrel{\hat{=}} \tau F(v)$, where $F(v)$ is a strictly positive, smooth function of $v$.

While the boundary value of $\zeta$ is fully specified, its transverse derivative ${\cal W}$ remains unconstrained. We fix this {extension freedom} by demanding that the evolution preserves the metric strictly in the normal plane:
\begin{equation}
\hspace*{-2mm} l^\mu n^\nu\mathcal L_\zeta g_{\mu\nu}
 \mathrel{\hat{=}} - \delta_\zeta (\phi + \psi)
 \mathrel{\hat{=}}{\cal W}-\kappa_\zeta=0,
 \ \ \ {\cal K}=2{\cal W}=2\kappa_\zeta.
 \label{eq:elf-boost-extension}
\end{equation}
This prescribes one transverse derivative of the generator. The projection on $\Sigma$, $q_\mu{}^\rho q_\nu{}^\sigma\mathcal L_\zeta g_{\rho\sigma} \mathrel{\hat{=}}2f\Theta^l_{\mu\nu}$, is left unconstrained,  allowing for arbitrary boundary expansion and shear. 

Together, \eqref{eq:elf-admissible-clock} and \eqref{eq:elf-boost-extension} dictate the normal transport:
\begin{equation}
 \mathcal L_\zeta l_\mu
   \mathrel{\hat{=}}-(\mathcal L_l f)l_\mu,
 \qquad
 \mathcal L_\zeta n_\mu
   \mathrel{\hat{=}}(\mathcal L_l f)n_\mu.
 \label{eq:elf-normal-transport}
\end{equation}
We fix the remaining clock reparametrization such that $\kappa_\zeta>0$ on a sufficiently small, smooth patch, and impose \eqref{eq:elf-boost-extension} relative to that clock. A general null geometry strictly prefers neither a unique clock nor a unique normalization.

The transverse extension~\eqref{eq:elf-boost-extension} directly equips $\zeta$ with a local physical observer interpretation. Since  $\zeta^2=-2fZ$, its normal derivative evaluates to
\begin{equation}
 n^\mu\nabla_\mu(\zeta^2)\big|_{\mathcal N}
   =2f{\cal W}=2f\kappa_\zeta>0.
 \label{eq:elf-timelike-side}
\end{equation}
Consequently, immediately off the boundary along $-n$, cf. Fig.~\ref{fig:geometric_setup}, the evolution vector becomes strictly timelike. There, it defines a canonical local observer congruence:
\begin{equation}
 V:=\sqrt{-\zeta^2},
 \qquad t^\mu:=\frac{\zeta^\mu}{V},
 \qquad t^\mu t_\mu=-1.
 \label{eq:elf-observers}
\end{equation}

We associate this physical clock with the thermodynamic temperature parameter
\begin{equation}
 \mathcal T_\zeta:=\frac{\kappa_\zeta}{2\pi}.
 \label{eq:elf-temperature}
\end{equation}
Local boosts in the $l,n$ plane normal to $\Sigma$ only act on the null frame \eqref{boost-l,n}. By augmenting this gauge transformation with the clock scaling \eqref{boost-f,Z}, the lapse $f$ effectively acts as a dressing field that absorbs the anomalous transformation of the non-affinity. As discussed below \eqref{eq:elf-admissible-clock}, this dressing factor is setting the local clock, and $\mathcal T_\zeta$ is measured in the units fixed by this clock. Consequently, $\mathcal{T}_\zeta$ represents a rigorously boost-invariant, ``dressed'' temperature. The above construction defines a set of physical observers locally around any point on $\Sigma$ or ${\cal N}$ and a strictly positive temperature. As we will see in the next section, $\mathcal T_\zeta$ acts as the temperature that is thermodynamically conjugate to the {entropy variation} appearing in the first law. This assignment does not by itself determine the response of a detector following $t^\mu$.

\section{The local first law}
\label{subsec:observer-energy}

Having specified the time (via $\zeta^\mu$ and its derivative), the charge variation associated with it, $\slashed{\delta} H_\zeta$ may be viewed as the physical energy variation conjugate to this evolution parameter. (The charge tied directly to the unit field $t^\mu$ \eqref{eq:elf-observers} would necessitate further normalization prescriptions, as surface charges are sensitive to derivatives of their generators.)

In direct analogy with the Kerr black hole—whose local horizon generator is a combination of asymptotic time translation and rotation, $\partial_t + \Omega \partial_\phi$—we decompose the null boundary generator $\zeta^\mu$ into a {chosen} time-evolution vector $\xi^\mu$ and a purely spacelike shift $\chi^\mu$ tangent to $\Sigma$:
\begin{equation}
 \zeta^\mu = \xi^\mu - \chi^\mu, \qquad l_\mu\chi^\mu = 0 = n_\mu \chi^\mu,
\end{equation}
and hence $\xi^2 = -2fZ + \chi^2$. At the null boundary ($Z \mathrel{\hat{=}} 0$), $\xi^\mu$ is generically spacelike ($\xi^2 \geq 0$). The term $-2fZ$ is negative immediately off the boundary along $-n^\mu$. If a bulk extension admits a locus where $\xi^2=0$ and a region where $\xi^2<0$, the Kerr analogy suggests existence of an ergoregion and timelike observers there. Their existence, and any asymptotic interpretation of $\xi^\mu$, require additional assumptions and are not used in the local first law we discuss here.

It is therefore physically motivated to define the macroscopic energy variation as the charge variation associated with the evolution vector $\xi$:
\begin{equation}
 \slashed{\delta} \text{E}:=\slashed{\delta} H_\xi.
 \label{eq:elf-shifted-energy}
\end{equation}
{Here $\slashed{\delta}\text{E}$ is the total gravity--matter surface-charge variation; its gravitational contribution is $\slashed{\delta}\text{E}-\slashed{\delta}H_\xi^{\mathrm m}$.}
The spatial shift $\chi^\mu$ then inherently captures the degrees of freedom tangential to $\Sigma$, including rotations. Denoting the associated {gravitational tangential-momentum charge density (angular momentum for rotational generators)} by $\text{\bf J}_\mu$, we have
\begin{equation}
 \text{\bf J}_\mu
 :=\frac{\boldsymbol\epsilon_\Sigma}{16\pi G}J_\mu.
 \label{eq:elf-momentum-density}
\end{equation}
By leveraging the trace--shear decomposition~\eqref{eq:ngc-deformation}, the complete surface-charge variation~\eqref{eq:ngc-full-charge} can be elegantly reorganized to isolate the distinct thermodynamic work channels:
\begin{align}
 \slashed{\delta} \text{E} {- \slashed{\delta}H_\xi^{\mathrm m}}- \int_\Sigma\chi^\mu\delta{\text{\bf J}}_\mu
 =&\frac{1}{16\pi G}\int_\Sigma
 \Big[2\kappa_\zeta\,\delta\boldsymbol\epsilon_\Sigma
       -2f\boldsymbol\epsilon_\Sigma\mathfrak D_\delta\theta_l
 \nonumber\\
      &-\frac{2f\theta_l}{d-2}\delta\boldsymbol\epsilon_\Sigma
       -f\boldsymbol\epsilon_\Sigma N_l^{\mu\nu}\delta q_{\mu\nu}
 \Big]\,.
 \label{eq:elf-energy-variation}
\end{align}

\noindent\textbf{Heat, angular-momentum and gravitational news work.}
\label{subsec:thermodynamic-balance}
To establish a coherent thermodynamic framework, we decompose the terms in \eqref{eq:elf-energy-variation} into heat and different work terms. This decomposition is not uniquely fixed within the covariant phase space formalism; however, one can readily recognize three kinds of terms in gravitational contributions (after subtracting the matter surface-charge contribution) in \eqref{eq:elf-energy-variation}: the angular-momentum work term, the energy flow associated with gravitational news/wave, and the remaining {gravitational terms}, which {play} the role of the heat.  

This yields the {gravitational} entropy one-form density:
\begin{align}
 \inbox{\slashed{\delta}\text{\bf s}_\zeta
 :=\frac{1}{4G}\Bigg[
 \delta\boldsymbol\epsilon_\Sigma-\frac{f}{\kappa_\zeta}
 \left(\boldsymbol\epsilon_\Sigma\mathfrak D_\delta\theta_l
       +\frac{\theta_l}{d-2}\delta\boldsymbol\epsilon_\Sigma\right)
 \Bigg].}
 \label{eq:elf-entropy-one-form}
\end{align}
In a strictly non-expanding sector ($\theta_l=0$, $\delta\theta_l=0$), it elegantly recovers the standard Bekenstein--Hawking variation $\slashed{\delta}\text{\bf s}_\zeta =\delta\boldsymbol\epsilon_\Sigma/(4G)$. {More general exactly integrable sectors, together with their
compatibility conditions for the prescribed clock, are characterized in Appendix~\ref{app:entropy-integrability}.}

With this assignment, \eqref{eq:elf-energy-variation} may be written as the familiar first-law:
\begin{align}
 \inbox{\hspace{-.1 cm}\begin{aligned}
 \slashed{\delta} \text{E}
 ={}&\int_\Sigma\mathcal T_\zeta\,
                       \slashed{\delta}\text{\bf s}_\zeta
     +\int_\Sigma\chi^\mu\delta\text{\bf J}_\mu+\slashed{\delta} \text{W}_{\mathrm{news}}
       + \slashed{\delta}H_\xi^{\mathrm m},
 \end{aligned}\hspace{-.1 cm}}
 \label{eq:elf-first-law}
\end{align}
where 
\begin{equation}
 \slashed{\delta} \text{W}_{\mathrm{news}}
 :=-\frac{1}{16\pi G}\int_\Sigma
       f\boldsymbol\epsilon_\Sigma N_l^{\mu\nu}\delta q_{\mu\nu},
 \label{eq:elf-news-work}
\end{equation}
denotes the gravitational news work. {The matter term $\slashed{\delta}H_\xi^{\mathrm m}$ includes all matter contributions, including those associated with the tangential part of $\xi$; $\text{\bf J}_\mu$ remains gravitational. It is not assumed to be pure work or to equal the null matter stress flux.} We close with some comments:
\begin{enumerate}[leftmargin=3mm, itemsep=0mm]
    \item $\mathcal{T}_\zeta, \slashed{\delta}\text{\bf s}_\zeta; \chi^\mu, \text{\bf J}_\mu$ and $\slashed{\delta} \text{W}_{\mathrm{news}}${, together with $\slashed{\delta}H_\xi^{\mathrm m}$}, and hence $\slashed{\delta} \text{E}$ and the first law, are all manifestly invariant under local boosts \eqref{boost-l,n} and \eqref{boost-f,Z}.
    \item $\slashed{\delta}\text{\bf s}_\zeta, \slashed{\delta} \text{W}_{\mathrm{news}}$ and $\slashed{\delta} \text{E}$ are one-forms over the solution space and are generically nonintegrable. 
    \item Even though $\slashed{\delta}\text{\bf s}_\zeta$ is not integrable for generic non-stationary backgrounds, {its evaluation along the prescribed time direction and the clock defines an entropy-production rate satisfying the second law}, as we show in the next part.
    \item The entropy one-form \eqref{eq:elf-entropy-one-form} depends explicitly on the physical clock $\zeta^\mu$, establishing that this dynamical gravitational entropy is a fundamentally observer-dependent quantity. Thus, the local entropy production rate derived in the following section is inextricably tied to this specific choice of observer, a central feature of gravitational thermodynamics recently emphasized in \cite{Shajiee:2026coz, Shajiee:2026cwg, Ciambelli:2023mir}.
    \item In \cite{Shajiee:2026coz}, we proposed a dynamical entropy formula associated with a null boundary. This dynamical entropy is the {gravitational} Noether charge associated with a null vector field generating the boundary  in a setting very closely related to our analyses here, and may be viewed as the integrable part of the entropy variation \eqref{eq:elf-entropy-one-form}.
\end{enumerate}

\vspace*{-2mm}
\section{The local second law}
\label{sec:entropy-second-law}
\vspace*{-2mm}

The entropy density defined in \eqref{eq:elf-entropy-one-form} is a manifestly boost-invariant $d-2$ form in spacetime and a one-form in the solution space. This is not defining an integrable charge, because it immediately follows from \eqref{eq:elf-entropy-one-form} that $\delta(\slashed{\delta}\text{\bf s}_\zeta) \neq 0$. {Appendix~\ref{app:entropy-integrability} gives the exact closure
condition and classifies the pointwise integrable sectors.} Nevertheless, the local second law established below does not require integrability.

The crux of the local second law lies in evaluating the entropy one-form {$(\slashed{\delta}\text{\bf s}_\zeta)[\delta]$} along the physical time evolution, $\delta_\zeta=\mathcal L_\zeta$, which actively transports the metric{, the matter fields,} and both normals along the flow. 
For a given $\Sigma$, \eqref{eq:ngc-area} and \eqref{eq:elf-normal-transport} yield
\begin{align}
 \delta_\zeta\boldsymbol\epsilon_\Sigma
   &=f\theta_l\boldsymbol\epsilon_\Sigma,\qquad b[\delta_\zeta]=-2\mathcal L_l f,
 \nonumber\\
 \delta_\zeta\theta_l
   &=f\mathcal L_l\theta_l,
 \qquad \mathfrak D_{\delta_\zeta}\theta_l = f\mathcal L_l\theta_l+\theta_l\mathcal L_l f.
 \label{eq:esl-geometric-evolution}
\end{align}

\textbf{Entropy-production.} One may define the entropy production density $\boldsymbol{\mathcal P}_\zeta$, as
\begin{equation}
\begin{split} \boldsymbol{\mathcal P}_\zeta:&=(\slashed{\delta}\text{\bf s}_\zeta)[\delta_\zeta]\\
 &=\frac{f^2\boldsymbol\epsilon_\Sigma}{4G\kappa_\zeta}
 \left(\kappa_l\theta_l-\mathcal L_l\theta_l
                         -\frac{\theta_l^2}{d-2}\right),
 \label{eq:esl-production-intermediate}
\end{split}\end{equation}
where we used $\kappa_\zeta=f\kappa_l+\mathcal L_l f$. Invoking the exact, non-affine Raychaudhuri equation for twist-free generators,
\begin{equation}
 \mathcal L_l\theta_l
 =\kappa_l\theta_l-\frac{\theta_l^2}{d-2}
   -N_l^{\mu\nu}N^l_{\mu\nu}
   -R_{ll},\quad R_{ll}:= R_{\mu\nu}l^\mu l^\nu,
 \label{eq:esl-raychaudhuri}
\end{equation}
we obtain
\begin{equation}
 \inbox{
 \boldsymbol{\mathcal P}_\zeta
 =\frac{f^2\boldsymbol\epsilon_\Sigma}{4G\kappa_\zeta}
                  \left(N_l^{\mu\nu}N^l_{\mu\nu}+ R_{ll}\right).}
 \label{eq:esl-local-second-law}
\end{equation}
{This is the production of the gravitational entropy defined by the chosen heat prescription; no matter entropy has been added to it.} 

{Upon the Einstein field equations, which yield $R_{ll} = 8\pi G T_{\mu\nu} l^\mu l^\nu$ for any cosmological constant, and assuming the matter energy-momentum tensor satisfies the null energy condition (NEC), ensuring $T_{\mu\nu}l^\mu l^\nu \geq 0$, we learn $R_{ll}\geq 0$. Furthermore, since $N_l^{\mu\nu}$ is a symmetric tensor with both legs along $\Sigma$,  $N_l^{\mu\nu}N^l_{\mu\nu}\geq 0$. Consequently, provided the NEC holds, $\boldsymbol{\mathcal P}_\zeta \geq 0$. This rigorously establishes an exact local second law: a strictly non-negative, pointwise rate of entropy production.} Remarkably, this exact local second law holds wherever the foliation is smooth, and the chosen clock satisfies $\kappa_\zeta>0$. It requires no perturbative expansion, no sign restriction on the expansion $\theta_l$, and no future endpoint conditions (such as a teleological event horizon). 

The entropy production comes from two sources: (i) the matter energy-momentum tensor (projected along the null vector field $l^\mu$); (ii) the shear squared, which is related to the energy density of the gravitational waves passing through the boundary ${\cal N}$. To see the latter more explicitly, let us study the news work along $\delta_\zeta$:
\begin{equation} \label{eq:esl-heat-work-check}
\begin{split}
 (\slashed{\delta} \text{W}_{\mathrm{news}})[\delta_\zeta]
 &=-\frac{1}{8\pi G}\int_\Sigma
          f^2\boldsymbol\epsilon_\Sigma N_l^{\mu\nu}N^l_{\mu\nu}\\ 
 & =-\int_\Sigma \bigg( \mathcal T_\zeta\boldsymbol{\mathcal P}_\zeta - \frac{ \boldsymbol\epsilon_\Sigma }{8\pi G}R_{\zeta\zeta}\bigg),
\end{split}
\end{equation}
where we used the fact that  ${q_\mu{}^\rho q_\nu{}^\sigma\delta_\zeta q_{\rho\sigma}}\mathrel{\hat{=}}2f\Theta^l_{\mu\nu}$ and that $f^2R_{ll}  \mathrel{\hat{=}} R_{\zeta\zeta}$.

Upon \eqref{eq:esl-heat-work-check}, the first law~\eqref{eq:elf-first-law} takes the form
\begin{equation}\label{deltaE-zeta}
\begin{aligned}
\slashed{\delta} \text{E}[\delta_\zeta]
=\int_\Sigma \bigg(\chi^\mu\delta_\zeta {\text{\bf J}}_\mu
 +\frac{\boldsymbol\epsilon_\Sigma}{8\pi G}R_{\zeta\zeta}\bigg)+ \big(\slashed{\delta}H_\xi^{\mathrm m}\big)[\delta_\zeta].
\end{aligned}
\end{equation}
Recalling \eqref{eq:elf-shifted-energy}, the left-hand-side of \eqref{deltaE-zeta} vanishes for $\chi=0$ ($\xi=\zeta$) case. Since $\slashed{\delta}H_\xi^{\mathrm m}$ is linear in $\xi$, $\slashed{\delta}H_\xi^{\mathrm m}[\delta_\zeta]= \slashed{\delta}H_\chi^{\mathrm m}[\delta_\zeta]+ \slashed{\delta}H_\zeta^{\mathrm m}[\delta_\zeta]$ and therefore, 
\begin{equation}
    \int_\Sigma \frac{\boldsymbol\epsilon_\Sigma}{8\pi G}R_{\zeta\zeta}+ \big(\slashed{\delta}H_\zeta^{\mathrm m}\big)[\delta_\zeta] =0.
\end{equation}
One can hence rewrite \eqref{eq:esl-heat-work-check} and \eqref{deltaE-zeta} as,
\begin{subequations} 
\begin{align}
 -\big(\slashed{\delta}H_\zeta^{\mathrm m}\big)[\delta_\zeta]&= \int_\Sigma  \mathcal T_\zeta\boldsymbol{\mathcal P}_\zeta+ (\slashed{\delta} \text{W}_{\mathrm{news}})[\delta_\zeta], \label{deltaGW-work-heat-heat}\\
\slashed{\delta} \text{E}[\delta_\zeta]
&=\int_\Sigma \chi^\mu\delta_\zeta {\text{\bf J}}_\mu+ \big(\slashed{\delta}H_\chi^{\mathrm m}\big)[\delta_\zeta].\label{deltaE-chi-chi}
\end{align}
\end{subequations}
$\big(\slashed{\delta}H_\zeta^{\mathrm m}\big)[\delta_\zeta]$ and $\mathcal T_\zeta\boldsymbol{\mathcal P}_\zeta$ may be respectively viewed as the heat flux of the matter and gravitational sectors, so \eqref{deltaGW-work-heat-heat} implies that the gravitational work (due to passage of gravitational waves through the boundary ${\cal N}$) plus the gravitational heat flux plus the heat flux of the matter sector vanishes. $\big(\slashed{\delta}H_\chi^{\mathrm m}\big)[\delta_\zeta]$ may be viewed as matter's spin rotational (kinetic) energy and hence $\slashed{\delta} \text{E}[\delta_\zeta]$ is the sum of matter and gravitational angular energies.

{In summary, the entropy production along the physical time evolution $\zeta^\mu$ is strictly non-negative. Locally, this production rate is exactly equivalent to the total dissipative energy flux—comprising {minus} the gravitational news work and the matter {stress-energy flux} traversing the boundary—divided by the local temperature $\mathcal{T}_\zeta$. Thermodynamic irreversibility and the local second law are thus inextricably governed by the purely dissipative nature of gravitational shear and the positivity of the null matter flux.}
{The stress-energy flux in this geometric production law and the matter surface-charge one-form $\slashed{\delta}H_\xi^{\mathrm m}$ are distinct quantities. The latter contributes to the total energy balance above and is not counted as an additional term in $\boldsymbol{\mathcal P}_\zeta$.}

\vskip 2mm

\textbf{History dependence and the scope of the second law.}
\label{subsec:entropy-interpretation}
As a consequence of the non-integrability of the entropy variation, the accumulated entropy fundamentally depends on the evolutionary history. Fixing the clock and normal transport, we define the accumulated entropy production along a phase-space history ${\gamma_t}$ (with tangent ${\dot\gamma_t}=\delta_\zeta$) by
\begin{align}
S_{\mathrm{hist}}[\gamma; {t_2}]
       -S_{\mathrm{hist}}[\gamma; {t_1}]
 :=\int_{{t_1}}^{{t_2}}\mathrm d{t}
       \int_\Sigma\boldsymbol{\mathcal P}_\zeta({\gamma_t})
       \geq0,
 \label{eq:esl-history-entropy}
\end{align}
where ${t}$ acts as the evolution parameter and a reference value fixes the additive constant, and  ${t_2\geq t_1}$. {The clock parameter obeys $\mathrm dt=\mathrm dv/F(v)$.} This yields a monotonically {non-decreasing}, history-dependent quantity. {Due to non-integrability of the entropy, different unrestricted paths in the solution space with identical endpoint configurations can give different entropy integrals. The monotonicity statement above applies to the prescribed physical evolution, rather than to arbitrary paths.}

This marks a stark departure from standard area-entropy paradigms, a fact already visible in vacuum (absence of matter fields). The area evolution,
\begin{equation}
 \mathcal L_\zeta\left(\frac{\boldsymbol\epsilon_\Sigma}{4G}\right)
 =\frac{f\theta_l}{4G}\boldsymbol\epsilon_\Sigma,
 \label{eq:esl-area-evolution}
\end{equation}
possesses no definite sign on a generic null boundary. For instance, a smooth, future-directed converging light cone in Minkowski spacetime has $\theta_l<0$ (decreasing area), yet its shear and $\boldsymbol{\mathcal P}_\zeta$ identically vanish. Thus, the production law~\eqref{eq:esl-local-second-law} definitively untangles entropy production from area growth: {in vacuum, this entropy production is driven by the shear-squared rather than area change}, and production-free evolution need not imply spacetime stationarity.

Crucially, the second law established here applies specifically to the entropy variation one-form evaluated on the dynamically prescribed future evolution (equipped with its normal transport and positive non-affinity). It does not artificially impose a sign on arbitrary field-space variations or on evolutions possessing arbitrary tangential shifts. The generic non-integrability encountered here is a physical reflection of this canonical heat prescription, rather than a breakdown of the second law itself. Our analysis {suggests} that, in general
settings involving dynamical gravity (including gravitational waves), an entropy defined as an integrable surface charge cannot guarantee the second law without additional assumptions or restrictions. Further investigation and perhaps a proof  of the proposed suggestion is left for future work.

\vspace*{-2mm}

\section{Near-stationary limit }
\label{sec:hwz-limit}
\vspace*{-2mm}
\textbf{Stationary reference geometry and perturbative assumptions.}
\label{subsec:stationary-reference}
To connect our exact dynamical framework with the Hollands–Wald–Zhang (HWZ) perturbative construction around a stationary black hole horizon \cite{Hollands:2024vbe}, consider a smooth family of {coupled solutions $\Phi(\lambda)=(g(\lambda),\Psi(\lambda))$, with} $g(\lambda)=g_0+\lambda h+O(\lambda^2)$ whose reference background $g_0$ possesses a connected, non-extremal bifurcate Killing horizon. We denote the future branch of the horizon by $\mathcal H^+$, its bifurcation surface by $\mathcal B$, and its Killing generator by $\xi_H^\mu$, which has a fixed normalization and constant surface gravity $\kappa_0>0$. Introducing an affine generator $k^\mu$ and its affine parameter $v$, we demand
\begin{align}
 k^\nu\nabla^{(0)}_\nu k^\mu&=0,
 \qquad \mathcal L_k v=1,
 \qquad v|_{\mathcal B}=0,
 \nonumber\\
 \xi_H^\mu&=\kappa_0 v k^\mu
 \qquad\text{on }\mathcal H^+.
 \label{eq:hwz-reference-generator}
\end{align}
The affine label canonically parameterizes the generators without relying on a specific spacetime metric ansatz. We take the comparison cuts (codimension-2 surfaces $\Sigma$) to be the level sets of $v$. {The reference matter fields are also invariant under $\xi_H$ in the chosen local gauge; a combined diffeomorphism and internal gauge symmetry requires the corresponding gauge-charge contribution, see e.g. \cite{Grumiller:2022qhx, Hajian:2022lgy}.

We map the perturbed horizon and its constant $v$ cuts to the reference background using a horizon-preserving gauge, firmly fixing the affine generator $k$ and parameter $v$ throughout the perturbation family. Crucially, the perturbed geometries themselves need not be stationary. In a notation closer to the one we have developed in this paper, at the reference solution, we set our physical clock to $\zeta_0=\xi_H$, ensuring that
\begin{align}
 l_0&=k,
 \qquad f_0=\kappa_0 v,
 \qquad \kappa_{\zeta,0}=\kappa_0,
 \nonumber\\
 \theta_{l,0}&=0,
 \qquad N^l_{\mu\nu,0}=0.
 \label{eq:hwz-reference-data}
\end{align}
The canonical Killing extension identically satisfies \eqref{eq:elf-boost-extension}  with ${\mathcal W_0}=\kappa_0$. On the future branch ($v>0$, so $f_0>0$), the entropy possesses a smooth, regular limit to $\mathcal B$ even though the observer generator vanishes at the bifurcation surface itself.

The perturbations of the cut data naturally expand as
\begin{align}
 \boldsymbol\epsilon_\Sigma(\lambda)
   &=\boldsymbol\epsilon_0+\lambda\boldsymbol\epsilon_1
       +O(\lambda^2),
 \nonumber\\
 \theta_l(\lambda)&=\lambda\theta_1+O(\lambda^2),
 \qquad
 N^l_{\mu\nu}(\lambda)=\lambda N^{(1)}_{\mu\nu}+O(\lambda^2).
 \label{eq:hwz-perturbation-expansion}
\end{align}
Note that a vanishing background expansion ($\theta_{l,0}=0$) does not generically force its first variation $\theta_1$ to zero. That additional restriction will be stated explicitly when we compare the leading entropy production laws.

\vskip 2mm
\textbf{Integrable limit and comparison with HWZ dynamical entropy \cite{Hollands:2024vbe}.}
\label{subsec:hwz-comparison}
For the comparison, we evaluate our entropy variation equation \eqref{eq:elf-entropy-one-form} directly at that reference:
\begin{align}
 \slashed{\delta}\text{\bf s}_\zeta\bigg|_0
 &=\frac{1}{4G}\left[
       \delta\boldsymbol\epsilon_\Sigma\big|_0
       -v\boldsymbol\epsilon_\Sigma \delta\theta_k\big|_0
       \right]
 \nonumber\\
 &=\delta\!\left[
       \frac{\boldsymbol\epsilon_\Sigma}{4G}(1-v\theta_k)
       \right]\bigg|_0.
 \label{eq:hwz-first-variation-matching}
\end{align}
Because the background expansion vanishes, {the additional term $\delta(v \boldsymbol\epsilon_\Sigma) \theta_k \big|_0$ vanishes when going from the first line to the second.} So, the entropy variation becomes integrable {on the linearized phase space}, and our entropy variation  precisely reproduces {the variation of} the HWZ entropy functional for Einstein gravity~\cite{Hollands:2024vbe} {(see also \cite{Visser:2024pwz, Rignon-Bret:2023fjq})}
\begin{equation}
 S_{\mathrm{HWZ}}[\Sigma]
 =\frac{1}{4G}\int_\Sigma
       \boldsymbol\epsilon_\Sigma(1-v\theta_k),
 \label{eq:hwz-entropy}
\end{equation}
where its thermodynamical interpretation is strictly validated at the perturbative order controlled by the stationary reference background. While HWZ entropy formula works for near-stationary configurations, it has been shown that a similar dynamical entropy formula can be obtained as a Noether charge associated with a null vector field generating the generic boundary, arbitrarily away from stationarity \cite{Shajiee:2026coz}.

Consequently, the entropy variation is perfectly integrable on the linearized phase space when background coefficients are held fixed. Explicitly, to the first order in $\lambda$,
\begin{equation}
 S_{\mathrm{HWZ}}^{[1]}(\lambda)
 =\frac{A_0}{4G}
  +\frac{\lambda}{4G}\int_\Sigma
       \left(\boldsymbol\epsilon_1
                   -v\boldsymbol\epsilon_0\theta_1\right),
 \label{eq:hwz-linearized-primitive}
\end{equation}
where $A_0=\int_\Sigma\boldsymbol\epsilon_0$. Since the background shear vanishes, the news work supplies no first-order contribution, the first law to the first order in $\lambda$ takes the form
\begin{equation}
 \begin{aligned}
 {\slashed{\delta}}\text{E}\bigg|_0
 ={}&\frac{\kappa_0}{2\pi}\delta S_{\mathrm{HWZ}}\bigg|_0
  +\int_\Sigma\chi_0^\mu\delta\text{\bf J}_\mu\bigg|_0+\slashed{\delta}H_\xi^{\mathrm m}\big|_0.
 \end{aligned}
 \label{eq:hwz-linearized-first-law}
\end{equation}
{Integrability of the gravitational entropy at this order does not by itself imply integrability of the total energy charge. The matter term is absent when matter fields and their variations are omitted (vacuum Einstein gravity), or when the specified matter boundary conditions make this surface-charge contribution vanish.}

This exact matching applies strictly to first variations at the stationary reference solution; it does not imply integrability of the nonlinear one-form  over a finite neighborhood in phase space. To explicitly delineate this departure,  define the comparison density $\text{\bf s}_{\mathrm{int}}$:
\begin{equation}
 \text{\bf s}_{\mathrm{int}}
 :=\frac{\boldsymbol\epsilon_\Sigma}{4G}(1-\beta \theta_l),\qquad \beta:=f/\kappa_\zeta\,.
 \label{eq:hwz-comparison-density}
\end{equation}
Direct differentiation then exposes the exact discrepancy:
\begin{align}
 \slashed{\delta}\text{\bf s}_\zeta-\delta\text{\bf s}_{\mathrm{int}}
 =\frac{\theta_l}{4G}\Bigg[
 &\frac{(d-3)}{d-2} \beta \delta\boldsymbol\epsilon_\Sigma+\boldsymbol\epsilon_\Sigma\delta \beta 
   +\frac12\beta \boldsymbol\epsilon_\Sigma\ b[\delta]
 \Bigg].
 \label{eq:hwz-nonlinear-difference}
\end{align}
While all correction terms vanish safely at the stationary reference, along a generic near-stationary trajectory they actively inject quadratic-order corrections to the integrated comparison with $\text{\bf s}_{\mathrm{int}}$, a reflection of the nonintegrability of $\slashed{\delta}\text{\bf s}_\zeta$. {The conditions under which $\text{\bf s}_{\mathrm{int}}$ is the exact integrable entropy (at nonlinear order), and its distinction from first-order HWZ matching, are explored and discussed in Appendix~\ref{app:entropy-integrability}.}

For vacuum perturbations (when the matter energy-momentum tensor $T_{\mu\nu}$ is zero), the first-order HWZ entropy remains constant along the horizon; thermodynamic entropy production \eqref{eq:esl-local-second-law} begins at quadratic order. In our framework, the exact production law naturally expands as
\begin{equation}
 \boldsymbol{\mathcal P}_\zeta
 =\lambda^2\frac{\kappa_0 v^2}{4G}
       \boldsymbol\epsilon_0
       N^{(1)}_{\mu\nu}N_{(1)}^{\mu\nu}
       +O(\lambda^3),
 \label{eq:hwz-leading-vacuum-production}
\end{equation}
where $N^{(1)}$ indices are raised using the reference metric on $\Sigma$, $q_0{}_{\mu\nu}$. 

The HWZ framework further restricts to vacuum horizon perturbations satisfying $\theta_1=0$~\cite{Hollands:2024vbe}. Under this further restriction, the exact difference in \eqref{eq:hwz-nonlinear-difference} affects integrated entropy variations only at cubic and higher orders. Applying the affine Raychaudhuri equation to \eqref{eq:hwz-entropy} with $\delta^2:=\partial_\lambda^2\big|_{\lambda=0}$ yields
\begin{equation}
 \frac{\mathrm d}{\mathrm dv}\delta^2 S_{\mathrm{HWZ}}
 =\frac{1}{2G}\int_\Sigma
       v\boldsymbol\epsilon_0
       N^{(1)}_{\mu\nu}N_{(1)}^{\mu\nu}.
 \label{eq:hwz-second-order-flux}
\end{equation}
The right-hand side is precisely twice the {integral of the} quadratic coefficient of $\boldsymbol{\mathcal P}_\zeta/(\kappa_0 v)$ over $\Sigma$—accounting for the conversion from the clock rate to the affine rate and from a quadratic order to a genuine second variation. This perfectly reproduces the positive modified canonical-energy flux derived by HWZ. Thus, while the HWZ flux relies on a stationary background and specific perturbative assumptions, our exact production law~\eqref{eq:esl-local-second-law} demonstrates that {non-negative}, shear-driven entropy production is a fundamental property of fully dynamical null boundaries { in vacuum, for the prescribed admissible clock}.

\section{Discussion}
\label{sec:conclusion}

In this work, we have dismantled the assumption that gravitational thermodynamics requires a stationary background, a preferred equilibrium state, or a teleological event horizon. By analyzing the covariant surface charges of general null boundaries in Einstein gravity, we have established an exact local first law and a rigorous second law that remain valid in deeply dynamical, non-equilibrium regimes. The cornerstone of this framework is the identification of an admissible physical clock and its transverse extension, which naturally defines a local observer congruence and canonically partitions the energy variation into heat, {gravitational} tangential work, and the energy carried by the gravitational waves, with the matter surface-charge contribution retained separately. Thus, this work completes the analysis of earlier work ``who writes the second law of thermodynamics'' \cite{Shajiee:2026cwg}.

The most striking physical consequence of this construction is the exact local entropy production law \eqref{eq:esl-local-second-law}. Evaluated along the boundary's physical evolution, the production rate is strictly non-negative, which, for pure Einstein gravity, is driven by gravitational shear. Remarkably, this second law holds {under the NEC} for any smooth null foliation with positive clock non-affinity, profoundly decoupling thermodynamic irreversibility from the (Hawking) area theorems. Gravity produces entropy even when the boundary's cross-sectional area is actively shrinking. 

Furthermore, our results reveal that the entropy one-form is generically non-integrable. Rather than signaling a breakdown of the thermodynamic description, this non-closure reflects a fundamental physical reality: we are dealing with a far-from-equilibrium open thermodynamical system. The entropy accumulated from this entropy variation is generally history-dependent. It cannot be uniquely assigned as a state function of an individual geometric cut. Yet, this exact, path-dependent framework smoothly connects to the known equilibrium and near-stationary physics: By restricting to a stationary reference geometry and imposing the HWZ perturbative conditions, our dynamical one-form agrees at first order with the variation of the integrable HWZ dynamical entropy, successfully reproducing its second-order shear flux while revealing the broader, exact non-linear landscape from which it emerges.

The local first law \eqref{eq:elf-first-law} differs from the standard textbook first law in two crucial respects: neither the energy nor the entropy variation is an exact one-form on the solution phase space. As we have discussed above, this non-integrability is a direct consequence of dealing with an open, far-from-equilibrium system and disappears when restricting to (near-)stationary configurations. Appendix~\ref{app:entropy-integrability} provides the general integrability conditions for the entropy variation, while the integrability of the energy variation is discussed in \cite{Adami:2021kvx}.

The thermodynamic arrow of time emerges when this generic variation is evaluated specifically along the physical observer's clock ($\delta \to \mathcal{L}_\zeta$). Under this specific restriction, $\delta$ transitions from an abstract comparison of alternate histories to representing the genuine \textit{local time evolution} within a single spacetime. Evaluated along this flow, the phase-space identities reduce to the standard balance equations, cleanly revealing the {non-negative, shear- and matter-driven} entropy production ($\boldsymbol{\mathcal P}_\zeta \ge 0$) required by the second law.

This realization elegantly elevates the foundational paradigm that gravity itself is the thermodynamics of spacetime \cite{Jacobson:1995ab}, extending it from local stationary patches to the entire dynamical phase space. Looking forward, this local formulation opens several compelling avenues. In particular, the explicit identification of a local observer congruence invites an operational analysis of how a localized quantum detector responds to this exact heat flux. 

Finally, by proving the exact local second law, we have completed the formulation of null surface thermodynamics \cite{Adami:2021kvx} as a fundamentally local, shear- and matter-driven, and history-dependent phenomenon. Our framework therefore provides a vital stepping stone toward understanding the statistical mechanics of generic non-stationary spacetimes, and will hopefully shed light on the standing issues regarding black holes and the information problem.

\vskip 2mm
\begin{acknowledgments}

\textbf{Acknowledgment.} 
The VRS and MMShJ acknowledge the Iran National Science Foundation (INSF) Research Chair Grant No.~4045163, and VT acknowledges INSF Grant No. 40506209. We used AI (OpenAI ChatGPT Astra 6) to hone the arguments.
\end{acknowledgments}

\vspace*{-5mm}


\appendix

\section{Local integrability of the clock conditions \texorpdfstring{\eqref{eq:elf-admissible-clock}}{(clock condition)}}
\label{app:admissible-clock}
\vspace*{-2mm}

In Section~\ref{subsec:evolution-generator}, we imposed the geometric condition \eqref{eq:elf-admissible-clock} $\mathcal L_\zeta n_\mu \mathrel{\hat{=}} \delta_\zeta \psi\, n_\mu$ to ensure that our physical time evolution vector $\zeta^\mu = f l^\mu + Z n^\mu$ properly respects the null boundary frame. Because $Z \mathrel{\hat{=}} 0$ on $\mathcal{N}$, this requirement reduces to a set of differential equations for the lapse function $f$. We must verify that these equations are locally integrable for any smooth foliation of the boundary. A conceptually similar argument was also made in \cite{Shajiee:2026cwg, Shajiee:2025cxl}.

Let us project the clock condition along the three linearly independent physical directions: the null generator $l^\mu$, the transverse normal $n^\mu$, and the spacelike directions tangent to the cuts $\Sigma$. Using the normalization $l^\mu n_\mu = -1$, these projections on \eqref{eq:elf-admissible-clock} yield:
\begin{align}
 \delta_\zeta\psi &\mathrel{\hat{=}} \mathcal{L}_l f, 
 \label{eq:app-clock-l-projection}\\
 n^\mu \nabla_\mu \ln f &\mathrel{\hat{=}}   n^\mu \mathcal{L}_l n_\mu, 
 \label{eq:app-clock-n-projection}\\
 D_a \ln f &\mathrel{\hat{=}} A_a. 
 \label{eq:app-clock-cut-projection}
\end{align}
Here, $D_a$ represents the covariant derivative projected onto the spatial cut $\Sigma$, and $A_a := q_a{}^\nu \mathcal{L}_l n_\nu$. 

Equation~\eqref{eq:app-clock-l-projection} simply defines the normalization shift $\delta_\zeta \psi$, and \eqref{eq:app-clock-n-projection} dictates the transverse derivative of $f$ just off the boundary. The critical restriction lies in  \eqref{eq:app-clock-cut-projection}. For a solution to exist, the spatial vector field $A_a$ must be curl-free on the $(d-2)$-dimensional cut ($D_{[a} A_{b]} = 0$).

To see that this integrability condition is automatically satisfied, consider the physical foliation of the null boundary. Let the scalar $v$ be a time coordinate that labels the successive spatial cuts $\Sigma_v$ (as depicted in Fig.~\ref{fig:geometric_setup}), chosen such that it monotonically increases along the null generators.  $n_\mu$ is the normal covector to these cuts, so its {pullback} to the boundary must be proportional to the gradient of $v${; we choose a compatible transverse extension of $v$ so that the following equality also holds for the ambient covectors}:
\begin{equation}
 n_\mu \mathrel{\hat{=}} -\tau \nabla_\mu v,
 \label{eq:app-clock-cut-lapse}
\end{equation}
where $\tau$ is a strictly positive lapse function. Since $l^\mu$ is tangent to the boundary and normalized by $l^\mu n_\mu = -1$, we immediately find $\tau = 1 / (l^\mu \nabla_\mu v)$.

Evaluating the Lie derivative of $n_\mu$ along the boundary generators, we find that the spatial vector $A_a$ takes the form:
\begin{equation}
 A_a = q_a{}^\nu \mathcal{L}_l (-\tau \nabla_\nu v) = - \tau q_a{}^\nu \mathcal{L}_l ( \nabla_\nu v) = - \tau  q_a{}^\nu \nabla_\nu (\mathcal{L}_l v).
\end{equation}
Using $\tau = 1 / (\mathcal{L}_l v)$, one immediately get
\begin{equation}
   A_a =  D_a \ln \tau\, .
   \label{eq:app-clock-exact-form}
\end{equation}
$A_a$ is an exact spatial gradient and hence its curl trivially vanishes. (Equivalently, in terms of the H\'aj\'i\v{c}ek one-form $\omega_a := -q_a{}^\nu n^\rho \nabla_\nu l_\rho$ and the normal acceleration $\bar\eta_a := -q_a{}^\nu l^\rho \nabla_\rho n_\nu$, one finds $A_a = -(\omega_a + \bar\eta_a)$, confirming that this specific combination is exact).

We can now integrate the spatial equation $D_a \ln f = D_a \ln {\tau}$ across $\Sigma$ to obtain the general solution on the boundary:
\begin{equation}
 f \mathrel{\hat{=}} \tau F(v), \qquad F(v) > 0.
 \label{eq:app-clock-boundary-solution}
\end{equation}
Here, $F(v)$ is an arbitrary smooth function that depends exclusively on the time label $v$. Physically, $F(v)$ represents our thermodynamic freedom to globally reparametrize the clock, ensuring that the prescribed evolution advances all spatial points on a given slice uniformly to the next slice.

Finally, to extend the clock $f$ away from the boundary ${\cal N}$ into the spacetime bulk, we utilize the transverse condition \eqref{eq:app-clock-n-projection}. Let $\rho$ be a transverse {flow parameter} along $-n^\mu$ such that $\rho=0$ on $\mathcal{N}$. The first derivative off the boundary is easily satisfied by an exponential extension:
\begin{equation}
 f(\rho) = (\tau F) \exp({-}\rho \, n^\mu \mathcal{L}_l n_\mu).
 \label{eq:app-clock-transverse-solution}
\end{equation}
{Here $\tau F$ and $n^\mu\mathcal L_l n_\mu$ are evaluated at $\rho=0$ and held fixed along the transverse curves. The minus sign in the exponent follows from $n^\mu\nabla_\mu=-\partial_\rho$ at the boundary.}
The {first transverse derivative} of the second component, $Z$, is similarly fixed by our geometric normal-plane condition in \eqref{eq:elf-boost-extension}.

This constructive proof guarantees that for any well-defined smooth foliation of the null boundary, an admissible physical clock $\zeta^\mu$ always exists {locally}. Notably, at a fixed null frame, setting $f=1$ {on the boundary} is mathematically permissible only if the lapse $\tau$ is spatially constant across every cut $\Sigma_v$ for any arbitrary $v$. {This condition need not hold for a general foliation at fixed null frame, so a non-trivial lapse function $f$ must be allowed in the exact local first law.}

While an admissible clock always exists {locally}, it is by no mean unique. The admissibility conditions leave the strictly positive function $F(v)$ completely undetermined. Because the evolution vector advances the cut label via $\mathcal{L}_\zeta v \mathrel{\hat{=}} f \mathcal{L}_l v = (\tau F)(1/\tau) = F(v)$, the function $F(v)$ dictates the elapsed physical clock time between successive spatial slices. {Explicitly, $\mathrm dt=\mathrm dv/F(v)$.} It is crucial to recognize that this freedom is a physical choice, not a gauge artifact. Under a local null-frame boost ($l \mapsto e^\sigma l$, $n \mapsto e^{-\sigma} n$, $f \mapsto e^{-\sigma} f$, $\tau \mapsto e^{-\sigma} \tau$), the ratio $F = f/{\tau}$ is manifestly gauge-invariant.

Therefore, choosing a new function $\widetilde{F}(v) = a(v)F(v)$ (with $a > 0$) defines a genuinely distinct physical clock, $\widetilde{\zeta} \mathrel{\hat{=}} a(v)\zeta$. The non-affinity of this new clock transforms as $\kappa_{\widetilde{\zeta}} = a\kappa_\zeta + \mathcal{L}_\zeta a$. If the scaling $a$ is a global constant, the transformation is trivial—it merely rescales the thermodynamic temperature (a global change of units) while perfectly preserving the unit four-velocity $t^\mu$ of the bulk observers when the bulk extension is rescaled by the same constant. However, for a dynamically varying reparametrization ($\mathcal{L}_\zeta a \neq 0$), enforcing the normal-plane metric preservation condition (${\widetilde{\mathcal W}} = \kappa_{\widetilde{\zeta}}$) rigorously dictates a modified transverse extension into the bulk. This modification actively alters the spacetime trajectories of the nearby timelike observers. Consequently, the non-uniqueness of $F(v)$ mathematically encodes a fundamental physical feature: the freedom to select which local observer congruence records the boundary thermodynamics.

\section{Integrability of the gravitational entropy one-form}
\label{app:entropy-integrability}

We determine when the gravitational entropy one-form \eqref{eq:elf-entropy-one-form} admits an entropy state function. We assume $d>2$ and in particular, the zero-temperature limit, in which division by $\kappa_\zeta$ requires a separate limiting prescription, is not included.

\subsection{The exact closure condition}
\label{app:entropy-closure}

Using $\beta=f/\kappa_\zeta$, as in \eqref{eq:hwz-comparison-density}, the entropy one-form is
\begin{equation}
\begin{split}
 \slashed{\delta}\text{\bf s}_\zeta
 =\frac{1}{4G}\bigg[\delta\boldsymbol\epsilon_\Sigma
 -\beta\bigg(&\boldsymbol\epsilon_\Sigma\delta\theta_l
 +\frac{\theta_l}{d-2}\delta\boldsymbol\epsilon_\Sigma
 -\frac12\boldsymbol\epsilon_\Sigma\theta_l b\bigg)\bigg].
 \label{eq:app-entropy-one-form}
\end{split}
\end{equation}
Here and below $\delta$ denotes the exterior derivative on solution phase space, and wedges  exclusively appears between  solution-space one-forms. Thus $b=\delta(\phi-\psi)$ implies $\delta b=0$. Taking the exterior derivative gives
\begin{equation}
\begin{split}
 \delta\bigl(\slashed{\delta}\text{\bf s}_\zeta\bigr)
 =-\frac{1}{4G}\bigg[&
 \delta(\beta\boldsymbol\epsilon_\Sigma)
       \wedge\delta\theta_l+\frac{1}{d-2}\delta(\beta\theta_l)
       \wedge\delta\boldsymbol\epsilon_\Sigma
 \\
 &-\frac12\delta(\beta\boldsymbol\epsilon_\Sigma\theta_l)
       \wedge b\bigg].
 \label{eq:app-entropy-curvature}
\end{split}
\end{equation}
Local integrability  amounts to the vanishing of this two-form throughout the chosen sector. For commuting allowed variations, this means
\begin{equation}
\begin{split}
 &\delta_1\!\left[(\slashed{\delta}\text{\bf s}_\zeta)[\delta_2]\right]-\delta_2\!\left[(\slashed{\delta}\text{\bf s}_\zeta)[\delta_1]\right]=0.
 \label{eq:app-entropy-commuting-test}
\end{split}
\end{equation}
For noncommuting (field-dependent) variations one must additionally subtract $(\slashed{\delta}\text{\bf s}_\zeta)[[\delta_1,\delta_2]]$ \cite{Compere:2015knw}. These are conditions on every pair of tangent variations, rather than on the evaluation of the one-form along a single physical evolution.

\subsection{General local solution}
\label{app:entropy-general-solution}

To solve \eqref{eq:app-entropy-curvature}, introduce a fixed positive reference form $\boldsymbol\epsilon_\star$ on the identified cut and set
\begin{equation}
 \boldsymbol\epsilon_\Sigma
   =a_\Sigma\boldsymbol\epsilon_\star,
 \qquad \delta\boldsymbol\epsilon_\star=0,
 \qquad a_\Sigma>0.
 \label{eq:app-entropy-area-ratio}
\end{equation}
This defines a scalar area ratio without choosing coordinates or a metric ansatz. The reference form need not come from a stationary solution. Locally, write
\begin{equation}
 \varpi:=\frac12(\phi-\psi),
 \qquad b=2\delta\varpi,
 \label{eq:app-entropy-boost-potential}
\end{equation}
and define
\begin{equation}
 \begin{split}
 X&:=a_\Sigma^{1/(d-2)}e^{-\varpi}\theta_l,
 \\
 Y&:=a_\Sigma^{(d-3)/(d-2)}e^{\varpi}\beta.
 \label{eq:app-entropy-XY}
 \end{split}
\end{equation}
In particular, $Y>0$ for the clocks considered here. Upon the identity
\begin{equation}
 Y\delta X
 =\beta\left(
 a_\Sigma\delta\theta_l
 +\frac{\theta_l}{d-2}\delta a_\Sigma
 -\frac12a_\Sigma\theta_l b\right),
 \label{eq:app-entropy-factorization}
\end{equation}
the entropy one-form and its exterior derivative take the form
\begin{align}
 \slashed{\delta}\text{\bf s}_\zeta
 &=\frac{\boldsymbol\epsilon_\star}{4G}
       (\delta a_\Sigma-Y\delta X),
 \label{eq:app-entropy-normal-form}\\
 \delta\bigl(\slashed{\delta}\text{\bf s}_\zeta\bigr)
 &=-\frac{\boldsymbol\epsilon_\star}{4G}
       \delta Y\wedge\delta X.
 \label{eq:app-entropy-rank-condition}
\end{align}
Consequently, the complete pointwise closure condition is $\delta Y\wedge\delta X=0$.

On a regular patch where $\delta X\neq0$, the most general local solution is
\begin{equation}
 Y=\mathscr F(X;x),\qquad \mathscr F>0,
 \label{eq:app-entropy-general-graph}
\end{equation}
where $x$ denotes an arbitrary given point on $\Sigma$, not a new coordinate choice, and positivity is required on the realized range of $X$. To establish necessity, choose local solution-space coordinates $(X,z^I)$. Equation~\eqref{eq:app-entropy-rank-condition} sets every transverse derivative $\partial Y/\partial z^I$ to zero. Conversely, \eqref{eq:app-entropy-general-graph} gives $\delta Y=\partial_X\mathscr F\,\delta X$ and hence closure. The function $\mathscr F$, including any parameters entering it, is fixed under variations within the chosen sector; its only varying argument is $X$. Allowing its parameters to vary independently would require testing closure again.

In terms of the original quantities, the general regular solution is
\begin{equation}
 \inbox{\begin{aligned}
 \beta&=e^{-\varpi}a_\Sigma^{-\frac{d-3}{d-2}}\mathscr F(X;x),\\
 X&=a_\Sigma^{\frac1{d-2}}e^{-\varpi}\theta_l.
 \end{aligned}}
 \label{eq:app-entropy-general-beta}
\end{equation}
If $\mathscr F(X;x)=\partial_X\mathscr U(X;x)$, the integrable entropy is
\begin{equation}
\begin{gathered}
 \inbox{
 \text{\bf s}_\zeta
 =\frac{\boldsymbol\epsilon_\star}{4G}
       [a_\Sigma-\mathscr U(X;x)], \qquad \slashed{\delta}\text{\bf s}_\zeta=\delta\text{\bf s}_\zeta,}
\end{gathered}
 \label{eq:app-entropy-general-primitive}
\end{equation}
up to a density constant on each connected solution-space sector, which may be conveniently set to zero (e.g. a la Wald-Zoupas argument \cite{Wald:1999wa}). It is worth nothing that the density \eqref{eq:hwz-comparison-density} in terms of the new variables \eqref{eq:app-entropy-XY} is written as
\begin{equation}
 \text{\bf s}_{\mathrm{int}}
 =\frac{\boldsymbol\epsilon_\Sigma}{4G}(1-\beta\theta_l)
 =\frac{\boldsymbol\epsilon_\star}{4G}(a_\Sigma-XY).
 \label{eq:app-entropy-comparison}
\end{equation}

These results respect the null-frame boost symmetry. Under \eqref{boost-l,n} and \eqref{boost-f,Z}, one has $\theta_l\mapsto e^\sigma\theta_l$,
$\beta\mapsto e^{-\sigma}\beta$, and $\varpi\mapsto\varpi+\sigma$; hence $X$ and $Y$ are invariant. A field-independent change of $\boldsymbol\epsilon_\star$ or of the integration constant in $\varpi$ only reparametrizes $X$, $Y$, and $\mathscr F$. It leaves the original entropy one-form and its closure unchanged. Upon fixing the boost symmetry, one can set $\varpi=0$ or set it such that $X=\theta_l, \beta=a_\Sigma^{-1} \mathscr F(X;x)$.

Finally, we comment that \eqref{eq:app-entropy-general-graph} can be interpreted as a coordinate transformation in the solution phase space, $X \to Y$. Such a change of coordinates is referred to as a change of slicing; see \cite{Golshani:2026lvc} for more discussions.

\subsection{Important special cases and the HWZ limit}
\label{app:entropy-special-cases}

\textbf{Area entropy and non-expanding sectors.}
Since $Y>0$, \eqref{eq:app-entropy-normal-form} shows that the area variation is recovered if and only if
\begin{equation}
 \delta X=0
 \quad\Longleftrightarrow\quad
 \mathfrak D_\delta\theta_l
   +\frac{\theta_l}{d-2}\delta\ln a_\Sigma=0.
 \label{eq:app-entropy-area-sector}
\end{equation}
Equivalently, $\theta_l=X_0e^{\varpi}a_\Sigma^{-1/(d-2)}$, with $\delta X_0=0$. The non-expanding sector $\theta_l=0$, with all variations preserving $\delta\theta_l=0$, is the special choice $X_0=0$. Thus, vanishing expansion is sufficient but not necessary, even for the entropy proportional to area. These are solution-space restrictions; $X_0$ need not be constant from one spacetime point to another.

\vskip 2mm
\textbf{The comparison entropy $\text{\bf s}_{\mathrm{int}}$ \eqref{eq:hwz-comparison-density} as an integrable charge.} One may compare \eqref{eq:app-entropy-comparison} and \eqref{eq:app-entropy-general-primitive}:
\begin{equation}
 \slashed{\delta}\text{\bf s}_\zeta
 -\delta\text{\bf s}_{\mathrm{int}}
 =\frac{\boldsymbol\epsilon_\star}{4G}X\delta Y,
 \label{eq:app-entropy-comparison-difference}
\end{equation}
which is \eqref{eq:hwz-nonlinear-difference} in the variables \eqref{eq:app-entropy-XY}. In particular, $\delta Y=0$ yields this comparison density as an exact nonlinear one-form.  $\delta Y=0$ may be equivalently wrriten as, 
\begin{equation}
 \delta\beta+\frac12\beta b
 +\frac{d-3}{d-2}\beta\delta\ln a_\Sigma=0.
 \label{eq:app-entropy-constant-Y}
\end{equation}
Where $\theta_l\neq0$, this condition is also necessary for $\text{\bf s}_{\mathrm{int}}$ itself to be the primitive. It is stronger than general integrability, for which the primitive is instead \eqref{eq:app-entropy-general-primitive}.

At a stationary reference with $\theta_{l,0}=0$, one has $X|_0=0$, so \eqref{eq:app-entropy-comparison-difference} vanishes at that solution even when $\delta X|_0\neq0$. This explains the first-order matching with HWZ in \eqref{eq:hwz-first-variation-matching}, with $\beta_0=v$ on the future branch $v>0$ and fixed background coefficients. It does not imply nonlinear closure in a neighborhood: $\delta Y\wedge\delta X$ can remain nonzero at the reference. Accordingly, the integrability in the linearized limit (i.e. the HWZ dynamical entropy) and the exact integrable sectors classified here are distinct statements.

\vskip 2mm
\textbf{Fixed normalization and fixed $\beta$.}
On a sector with $b=0$ and $\delta\beta=0$, \eqref{eq:app-entropy-curvature} reduces to
\begin{equation}
 \delta\bigl(\slashed{\delta}\text{\bf s}_\zeta\bigr)
 =-\frac{\beta}{4G}\frac{d-3}{d-2}
       \delta\boldsymbol\epsilon_\Sigma\wedge\delta\theta_l.
 \label{eq:app-entropy-fixed-beta}
\end{equation}
For $d=3$, this vanishes identically, and $\text{\bf s}_{\mathrm{int}}$ is already the integrable entropy without further restrictions on area and expansion variations. This feature dovetails with the fact that in $3d$ Einstein gravity there are no news and surface charges are integrable \cite{Adami:2020ugu, Geiller:2021vpg}. For $d>3$, closure requires $\delta\boldsymbol\epsilon_\Sigma\wedge\delta\theta_l=0$. Where $\delta a_\Sigma\neq0$, this means $\theta_l=h(a_\Sigma;x)$ locally; constant-area and fixed-expansion sectors are also included. These statements for these examples remain subject to clock admissibility and the coupled field equations, discussed below.

\subsection{Compatibility with the physical clock}
\label{app:entropy-clock-compatibility}

The arbitrary function in \eqref{eq:app-entropy-general-beta} solves the entropy closure equation; it does not independently solve the clock equations. Appendix~\ref{app:admissible-clock} gives $f=\tau F(v)$ on each connected cut, where $\tau=(\mathcal L_l v)^{-1}$ and $F(v)>0$. Consequently,
\begin{equation}
 \beta^{-1}
 =\kappa_l+\mathcal L_l\ln\tau
      +\frac{1}{\tau}\partial_v\ln F.
 \label{eq:app-entropy-clock-beta}
\end{equation}
Define the boost-invariant scalar
\begin{equation}
 C:=\tau\left(
      \beta^{-1}-\kappa_l-\mathcal L_l\ln\tau\right).
 \label{eq:app-entropy-clock-C}
\end{equation}
If the clock has not yet been selected, a smooth positive candidate $\beta$ is locally compatible with a clock of this form precisely when $D_a C=0$, so that $C=C(v)$ on each connected cut. The boundary clock is then reconstructed by
\begin{equation}
 F(v)=F_0\exp\!\left(\int_{v_0}^{v}C(s)\,\mathrm ds\right),
 \qquad F_0>0.
 \label{eq:app-entropy-clock-reconstruction}
\end{equation}
It obeys $\kappa_\zeta=f/\beta>0$; its transverse derivatives are fixed as in Appendix~\ref{app:admissible-clock} and \eqref{eq:elf-boost-extension}. If a clock prescription $F$ is already fixed, the stronger requirement is $C=\partial_v\ln F$ for that prescription, with its induced solution-space variations. Thus, the admissible integrable sectors are the intersections of the closure conditions above with the coupled solution space and the prescribed clock conditions.

\subsection{Physical evolution and the second law}
\label{app:entropy-integrable-evolution}

Now, let us investigate the evolution of \eqref{eq:app-entropy-general-graph} along the observer $\zeta$,
\begin{equation}
 \delta_\zeta Y
 =\partial_X\mathscr F\,\delta_\zeta X,
 \label{eq:app-entropy-sector-preservation}
\end{equation}
with the clock prescription and all fixed auxiliary data treated consistently. One must not independently replace the induced variation of a field-dependent clock by an assumed scalar transport law. Along a solution-space history with tangent $\delta_\zeta$, exactness gives
\begin{equation}
 \delta_\zeta\text{\bf s}_\zeta
 =\frac{f^2\boldsymbol\epsilon_\Sigma}{4G\kappa_\zeta}
   \left(N_l^{\mu\nu}N^l_{\mu\nu}
                +R_{\mu\nu}l^\mu l^\nu\right)\geq0,
 \label{eq:app-entropy-integrable-second-law}
\end{equation}
where the inequality uses the NEC. The derivative here is the solution-space derivative with the same fixed-cut identification as the main text. No independent transport of the reference form or of the functions defining the sector is implicit.

This tangency requirement has a concrete consequence for the area sectors. Equations~\eqref{eq:esl-geometric-evolution} and \eqref{eq:esl-raychaudhuri} yield
\begin{equation}
\begin{split}
 \delta_\zeta X
 =e^{-\varpi}a_\Sigma^{1/(d-2)}\big[&\kappa_\zeta\theta_l
-f\big(N_l^{\mu\nu}N^l_{\mu\nu}
              +R_{\mu\nu}l^\mu l^\nu\big)\big].
 \label{eq:app-entropy-X-evolution}
\end{split}
\end{equation}
Therefore preservation of $X=X_0$ requires
\begin{equation}
 \theta_l=\beta\left(N_l^{\mu\nu}N^l_{\mu\nu}
                   +R_{\mu\nu}l^\mu l^\nu\right)\geq0.
 \label{eq:app-entropy-area-sector-evolution}
\end{equation}
Algebraically integrable area sectors with negative expansion cannot also be preserved by this physical evolution under the NEC. For a preserved non-expanding sector, the shear and null matter flux both vanish. This is a local statement about the null boundary and does not establish spacetime stationarity. More generally, closure establishes an entropy state function; it establishes neither quasi-static evolution nor vanishing entropy production. Integrability of this gravitational entropy also does not imply integrability of the total energy or of the separately retained matter surface-charge contribution.

\subsection{Global exactness and integrated entropy}
\label{app:entropy-global-integrability}

Closure, recalling Poincar\'e lemma, is a local condition. A globally single-valued entropy on a connected sector additionally requires vanishing periods,
\begin{equation}
 \oint_\gamma\slashed{\delta}\text{\bf s}_\zeta=0
 \quad\text{for every closed loop on solution space}.
 \label{eq:app-entropy-periods}
\end{equation}
When $X$ and $Y$ are defined along the whole loop, this is equivalent to $\oint_\gamma Y\delta X=0$ at each point on $\Sigma$. The condition is automatic on a simply connected sector for a globally defined smooth closed one-form. In contrast, restriction to a single one-dimensional trajectory makes every two-form vanish automatically; integrating along one trajectory cannot establish integrability on a larger sector.

Our classification concerns the local entropy density in \eqref{eq:elf-entropy-one-form}. It supplies an integrable entropy $S_\zeta[\Sigma]=\int_\Sigma\text{\bf s}_\zeta$ when the integral and its variations exist. If one requires only integrability after integration over a fixed $\Sigma$, the necessary and sufficient local closure condition is instead
\begin{equation}
 \int_\Sigma\boldsymbol\epsilon_\star\,
                   \delta Y\wedge\delta X=0.
 \label{eq:app-entropy-integrated-closure}
\end{equation}
This can be weaker because of spatial cancellations or total derivatives, with the latter subject to the $\Sigma$ boundary conditions. No equivalence between integrated and pointwise closure is assumed. The general solution above is therefore a classification of pointwise integrable sectors for the specified heat and clock prescriptions, with the additional global and dynamical requirements stated explicitly.


\bibliographystyle{fullsort.bst}
\bibliography{reference}
\end{document}